\documentclass[aps,prl,superscriptaddress,amsmath,amssymb,aps,twocolumn,]{revtex4-2}

\usepackage{graphicx}% Include figure files
\usepackage{dcolumn}% Align table columns on decimal point
\usepackage{bm}% bold math
\usepackage{latexsym,epsfig}
\usepackage{graphicx}
\usepackage{verbatim}
\usepackage{comment}
\usepackage{amsmath}
\usepackage{amssymb}
\usepackage{stmaryrd}
\usepackage{color}
\usepackage{epstopdf}
\usepackage{grffile}
\usepackage[normalem]{ulem}
\usepackage{float}
\usepackage{tikz}
\usetikzlibrary{quantikz}
\DeclareGraphicsExtensions{.eps}

\newcommand{\beq}{\begin{equation}}
\newcommand{\eeq}{\end{equation}}
\newcommand{\bea}{\begin{eqnarray}}
\newcommand{\eea}{\end{eqnarray}}
\newcommand{\ben}{\begin{eqnarray*}}
\newcommand{\een}{\end{eqnarray*}}
\newcommand{\bfig}{\begin{figure}}
\newcommand{\efig}{\end{figure}}

\usepackage{hyperref}
\hypersetup{
    colorlinks=true,      
    urlcolor=blue,
    citecolor=blue,
    linkcolor=blue
}

\begin{document}
\title{Realizing non-trivial doublon formation using a quantum computer}
\author{Biswajit Paul}
\affiliation{School of Physical Sciences, National Institute of Science Education and Research, Jatni 752050, India}
\affiliation{Homi Bhabha National Institute, Training School Complex, Anushaktinagar, Mumbai 400094, India}

\author{Tapan Mishra}
\email{mishratapan@gmail.com}
\affiliation{School of Physical Sciences, National Institute of Science Education and Research, Jatni 752050, India}
\affiliation{Homi Bhabha National Institute, Training School Complex, Anushaktinagar, Mumbai 400094, India}

%\author{Biswajit Paul$^{1,2}$ and Tapan Mishra$^{1,2}$}
%\email{mishratapan@gmail.com}
%%$^1$School of Physical Sciences, National Institute of Science Education and Research, Jatni 752050, India\\
%$^2$Homi Bhabha National Institute, Training School Complex, Anushaktinagar, Mumbai 400094, India}

\date{\today}

\begin{abstract}
Dynamical formation of doublons or onsite repulsively bound pairs of particles on a lattice is a highly non-trivial phenomenon. In this work, we show the signatures of doublon formation in a quantum computer by simulating the continuous time quantum walk in the framework of the one dimensional extended Fermi-Hubbard model. By considering two up-component and one down-component particles initially created at the three neighbouring sites at the middle of the lattice and allowing intra- (inter-) component nearest neighbour (onsite) interactions we show the formation a stable onsite doublon in the quantum walk. The probability of such doublon formation is more (less) if the hopping strength of the down particle is weaker (stronger) compared to the up particle. 
On the contrary, for an initial doublon along with a free up particle, the stability of the doublon is more prominent than the doublon dissociation in the dynamics irrespective of the hopping asymmetry between the two components. 
We first numerically obtain the signatures of the stable doublon formation in the dynamics and then observe them using Noisy Intermediate-Scale Quantum (NISQ) devices.
\end{abstract}

\maketitle
{\em Introduction.-}
Dynamics of interacting many-body systems following a sudden quench of system parameters reveals fascinating phenomena in condensed matter~\cite{Bloch2008, Moeckel_2010, Polkovnikov2011}. Interactions along with particle statistics and lattice geometries are known to play crucial roles in establishing non-trivial scenarios in the dynamics which are otherwise absent in systems at equilibrium. 
One such phenomenon is the formation of doublons that are the repulsively bound onsite pairs of constituent particles formed due to strong inter-particle interaction.
Although the concept of doublons was originally discussed in the context of Hubbard model~\cite{Hubbard1963, Yang1989}, recent progress in the field of ultracold atoms in optical lattices have enabled the observation of stable doublons of bosonic particles ~\cite{Winkler2006}. Subsequently doublons have been observed in the context of several bosonic and fermionic Hubbard models~\cite{ Jordens2008, Niels2010, Greif2011, Nagerl2013, Jurgensen2014, Xia2015, Covey2016, hond2022} and their dynamical properties have been studied in recent years~\cite{petrosyan2007quantum, Khomeriki2010, Hofmann2012, Kolovsky_2012, BOLonghi2012, lea_Santos_2012, Chudnovskiy2012, doublon_Longhi2013, Boschi2014, Zakrzewski2017, maria_doublon}. 

While the study of controllable formation of doublons and their stability in many-body systems remain challenging both theoretically and experimentally, the quench dynamics of few interacting particles or the continuous time quantum walk (QW) offers a suitable platform to realise such phenomena in the simplest possible way~\cite{Kempe2003,Venegas-Andraca2012}. In this context, it has been well understood that a pair of particles (bosons or fermions of two different spins) when initialized on a particular site form a stable doublon in the QW due to strong onsite interaction. However, when initiated at the nearest neighbour (NN) sites, a doublon is not formed as strong repulsion between the constituents prohibits the overlap of their wavefunctions~\cite{Greiner_walk,lahini2012qw}. 
This question remained open until the recent prediction of stable onsite doublon formation in the QW of initially non-local bosons in the QW of two-component bosons~\cite{giri2022nontrivial}. 
However, the observation of such non-trivial doublon remains challenging due to the complexity of the model involved.

Recently, quantum computers have become very useful tools to simulate the dynamics of condensed matter systems and observe various physical phenomena~\cite{ Smith2019, Motta2020, Vovrosh2021, Sun2021, Kamakari2022, Chen2023, Koh2023, Hoke2023, skin_effect_Shen2023}. 
Several quantum computing simulations have been performed in the context of condensed matter physics using such NISQ devices which includes topological phases~\cite{Emanuele2020, Smith2022, Koh2022, topo_chiral_koh2022, Tan2023, HOTI_koh_2023}, Floquet systems~\cite{Mei2020,google_noisy_resilient, Harle2023}, Fermi-Hubbard models~\cite{arute2020,Martin2022, Stanisic2022}, spin systems~\cite{Smith2019, Sun2021, chen_AKLT, Wang2023, rosenberg2023dynamics}, quantum scar~\cite{Chen2022}, time crystal\cite{Mi2022, chen2023time_crystal}, non-hermitian physics~\cite{skin_effect_Shen2023} etc. However, the operations involving qubits or two-level systems in such quantum devices play a real bottleneck for the simulation of the dynamics of bosonic Hamiltonians involving finite short-range interactions and hence the observation of the non-trivial onsite doublons. To counter such difficulties, it is essential to consider systems of fermions whose dynamics can be simulated using the exiting quantum computers.

In this work, we predict the formation of a non-trivial doublon in the QW of two-component fermions ($\uparrow$ and $\downarrow$) and observe it in a quantum computing simulation. 
By considering an initial state of three nearest neighbour fermions in a one dimensional chain, we show that a stable onsite doublon ($\uparrow\downarrow$ pair) is formed due to the competition between the onsite inter-particle interaction and nearest-neighbour (NN) intra-particle interaction. We find that the formation, dissociation, and stability of such doublon strongly depend on the hopping asymmetry between the two components, interplay between interactions and the initial configuration considered. We first numerically establish the formation of this non-trivial doublon and then perform a digital quantum circuit implementation in the framework of the Fermi-Hubbard model on  NISQ devices.

{\em Model and approach.-}
The system of two-component fermions possessing onsite and NN interaction in one dimension is described by the extended Hubbard model which is given by the Hamiltonian 
\begin{equation}
    \begin{split}
        \hat{H} = \sum_{\substack{j \\ \sigma={\uparrow, \downarrow}}}-J_{\sigma}(\hat{a}_{j,\sigma}^\dagger \hat{a}_{j+1,\sigma}&+\text{H.c.})        +U_{\uparrow\downarrow}\sum_j\hat{n}_{j,\uparrow}\hat{n}_{j,\downarrow}\\
        &+\sum_{\substack{j \\ \sigma,\sigma'={\uparrow, \downarrow}}}V_{\sigma\sigma^{\prime}}\hat{n}_{j,\sigma}\hat{n}_{j+1,\sigma^{\prime}},
    \end{split}
    \label{eqn:ham}
\end{equation}
where ${\hat{a}^\dagger}_{j,\sigma}$$(\hat{a}_{j,\sigma})$ are the creation (annihilation) operators for the two types of fermions denoted by $\sigma=\uparrow, \downarrow$ and $\hat{n}_{j,\sigma}=\hat{a}^\dagger_{j,\sigma}\hat{a}_{j,\sigma}$ is the particle number at the $j^{th}$ lattice site. $J_\sigma$ is the NN hopping strength of the particles, $V_{\sigma\sigma^{\prime}}$ and $U_{\uparrow\downarrow}$ are the NN and inter-component onsite interaction strengths respectively. We define $\delta=\frac{J_{\downarrow}}{J_{\uparrow}}$ as the hopping imbalance between the up and down particles.
\begin{figure}[t!]
    \centering
    \includegraphics[width=1\columnwidth]{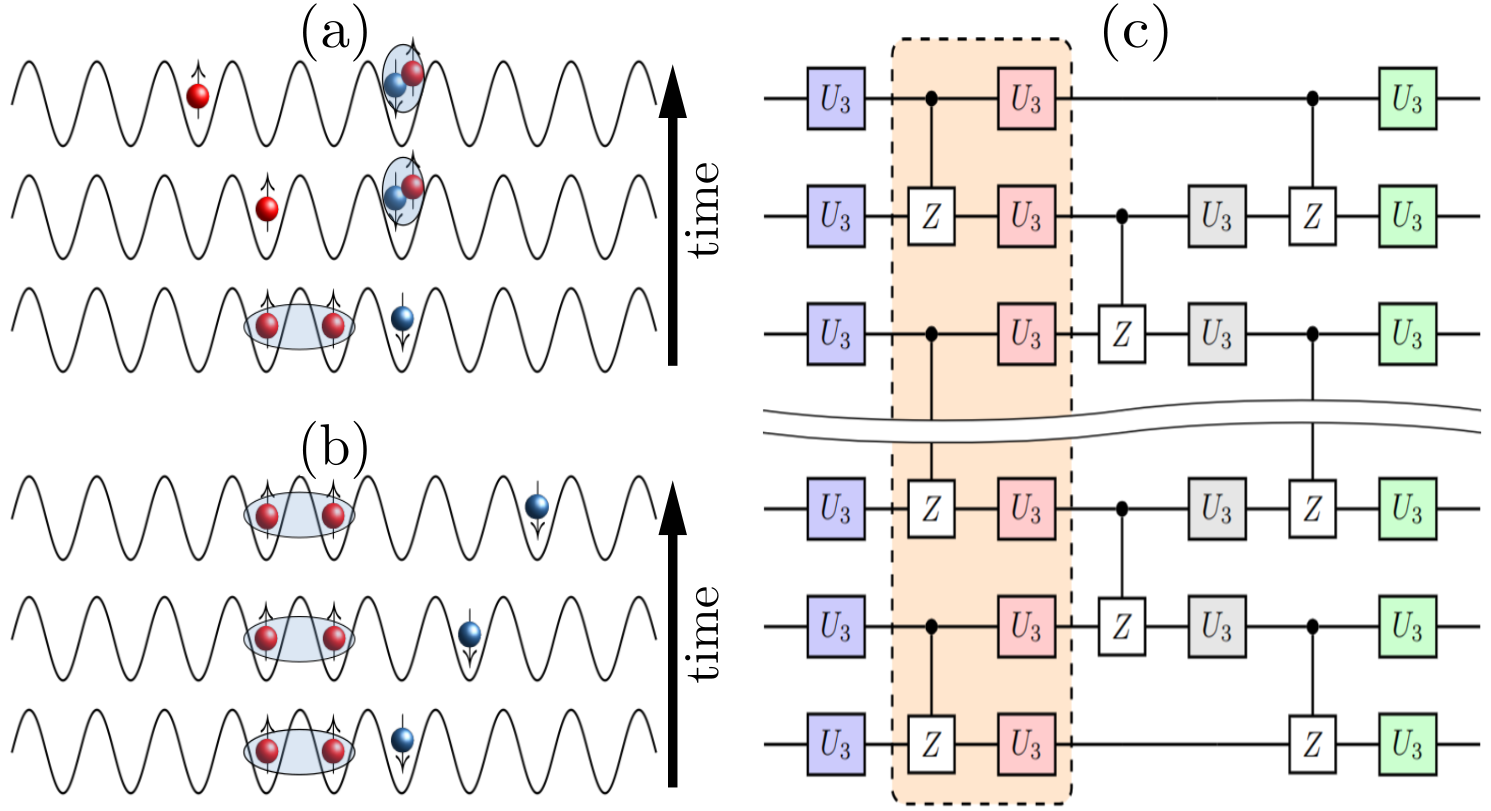}
    \caption{Shows the schematic for the short-time evolution of the initial state $|\psi(0)\rangle=\hat{a}_{-1,\uparrow}^{\dagger} \hat{a}_{0,\uparrow}^{\dagger}\hat{a}_{1,\downarrow}^{\dagger}|vac\rangle$. Two possible energy conserving time evolution processes are shown in (a) and (b). (c) shows the recompiled circuit, consisting of an alternative single-qubit $U_3$ layer and two-qubit control-Z layer. A gate round consists of a single layer of single-qubit gate and a single layer of two-qubit gate shown in the dashed box.}
    \label{fig:toy_plot}
\end{figure}

The quantum state at time $t\neq 0$ is determined from the solution of the time-dependent Schrodinger equation, $|\psi(t)\rangle=e^{-i\hat{H}t}|\psi(0)\rangle$, where $|\psi(0)\rangle$ is the initial state. Our numerical study is based on the time evolved block decimation (TEBD) ~\cite{Vidal2003, Vidal2004} approach using the OSMPS library~\cite{wall2015out, JASCHKE2018} by considering systems of size $L=51$ sites. 
For the quantum circuit implementation, we consider a limited number of qubits and further compare the results with the data obtained using the exact diagonalization (ED) method for an equivalent system. Although we restrict ourselves to smaller system sizes and short-time dynamics in our quantum computing simulation, we are able to capture important physics using certain kinds of error mitigation and circuit optimization techniques.

{\em Results.-} The QW is studied by considering an initial state of two $\uparrow$ and one $\downarrow$ particles located at the three consecutive sites such that the two $\uparrow$ particle are in the NN sites and the $\downarrow$ is on the adjacent site on the right of the right $\uparrow$ particle as depicted in Fig.~\ref{fig:toy_plot}(a) (bottom figure). The initial state corresponding to this configuration is given as $|\psi(0)\rangle=\hat{a}_{-1,\uparrow}^{\dagger} \hat{a}_{0,\uparrow}^{\dagger}\hat{a}_{1,\downarrow}^{\dagger}|vac\rangle$. 
For this choice of the initial state, although  $U_{\uparrow\downarrow}$ and both $V_{\uparrow\uparrow}$ and $V_{\uparrow\downarrow}$ terms are allowed, we assume $V_{\uparrow\downarrow}=0$ to capture the desired physics due to the competing interactions. From here onwards we denote $U_{\uparrow\downarrow}$ and $V_{\uparrow\uparrow}$ as $U$ and $V$ respectively. We consider a stronger NN interaction $V=10$ and vary $U$ to study their combined effect on the dynamics. When $U=0$, it is expected that the two $\uparrow$ particles form a repulsively bound NN pair ($\uparrow\uparrow$) at their initial position~\cite{bloch_magnon_expt, Luis-Arya} and move together whereas the down particle perform independent particle QW. However, as $U$ becomes finite and comparable with $V$ (i.e. $U\sim V$), we obtain finite probabilities of both onsite doublon along with an NN pair ($\uparrow\uparrow$) in the dynamics for equal hopping strengths of both the components (i.e. $\delta=1$). This behaviour can be quantified by comparing the probabilities of the doublon and the NN pair ($\uparrow\uparrow$) formation defined as 
\begin{equation}
    P_{\uparrow\downarrow}=\sum_i^{L} \langle \hat{n}_{i,\uparrow}\hat{n}_{i,\downarrow}\rangle \ \  \text{and} \ \  
    P_{\uparrow\uparrow}=\sum_i^{L-1} \langle \hat{n}_{i,\uparrow}\hat{n}_{i+1,\uparrow}\rangle
    \label{eqn:pair_prob}
\end{equation}
respectively.

\begin{figure}[t!]
   \centering
   \includegraphics[width=1\columnwidth]{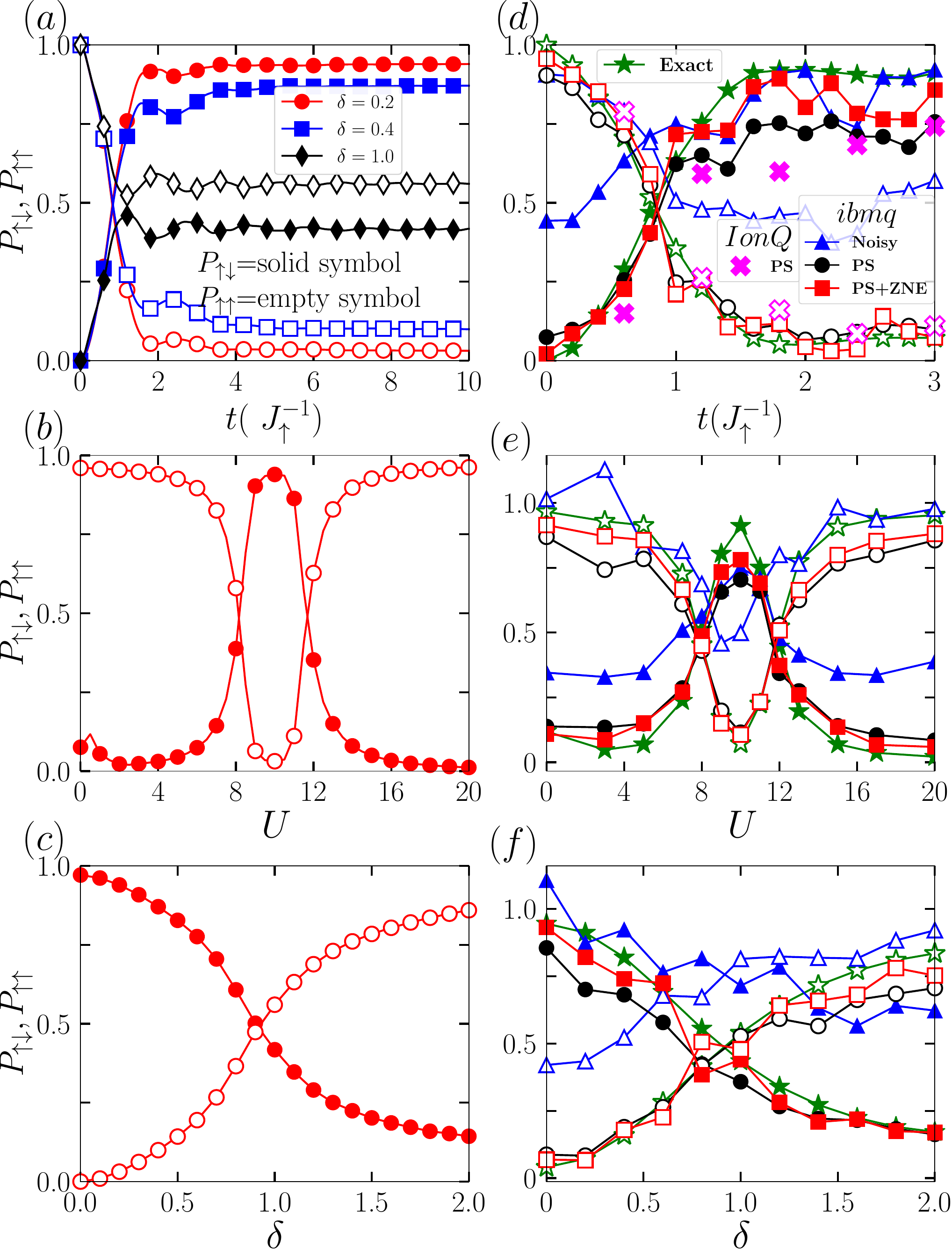}
   \caption{Left panel shows the results from the classical simulation with $L=51$. (a) $P_{\uparrow\downarrow}$ and $P_{\uparrow\uparrow}$  are plotted as a function of time at the resonance condition ($V = U=10$) for different values of hopping imbalance $\delta$. (b) $P_{\uparrow\downarrow}$ and $P_{\uparrow\uparrow}$ are plotted against $U$ for $\delta=0.2$ and $V=10$ at a particular time $t=10J_{\uparrow}^{-1}$. (c) $P_{\uparrow\downarrow}$ and $P_{\uparrow\uparrow}$ plotted as a function of $\delta$, for $U=V=10$ at time $t=10J_{\uparrow}^{-1}$. The right panel shows the results from the digital quantum simulation for $7$ lattice sites and short-time dynamics. (d) $P_{\uparrow\downarrow}$ and $P_{\downarrow\downarrow}$ as a function of time $(t)$ for $U=V = 10$ and $\delta=0.2$. (e) $P_{\uparrow\downarrow}$ and $P_{\uparrow\uparrow}$ at  time $t=1.6(J^{-1})$ as a function of $U$ for $\delta=0.2$. (f)  $P_{\uparrow\downarrow}$ and $P_{\uparrow\uparrow}$ at  time $t=1.6(J^{-1})$ as a function of $\delta$ for $U=10$. In both (e) and (f) we consider $V=10$. For all the figures $P_{\uparrow\downarrow}$ ($P_{\uparrow\uparrow}$) is denoted by solid (empty) symbols. In (d-f), the green stars, bule triangles, black circles and red squares denote the exact, noisy, PS and PS+ZNE data respectively obtained using the $ibmq\_brisbane$ hardware. The magenta crosses in (d) denote the PS mitigated data from the $IonQ$-$Aria~1$ hardware.}
   \label{fig:p_up_down}
\end{figure}

In Fig.~\ref{fig:p_up_down}(a), we plot the time evolution of $P_{\uparrow\downarrow}$ (filled diamonds) and $P_{\uparrow\uparrow}$ (empty diamonds) for $\delta=1$ and $U=V=10$ which saturate to finite values close to $0.5$. The almost equal probabilities for both the $\uparrow\downarrow$ and $\uparrow\uparrow$ bound pairs is due to the almost equal energies of both the states at $U=V$. However, introducing a finite hopping imbalance i.e. $\delta < 1$, $P_{\uparrow\downarrow}$ (filled symbols) clearly dominates over $P_{\downarrow\downarrow}$ (empty symbols) in the time evolution which are shown as blue squares for $\delta=0.4$ and red circles for $\delta=0.2$ respectively. This indicates that when the hopping imbalance is stronger (i.e. $\delta << 1$), the $\uparrow\uparrow$ pair tends to dissociate completely and a stable onsite doublon ($\uparrow\downarrow$) is formed in the QW. To quantify this behaviour further, we plot the saturated values of $P_{\uparrow\downarrow}$ (filled circles) and $P_{\uparrow\uparrow}$ (empty circles) at $t=10(J_{\uparrow}^{-1})$ in the time evolution as a function of $U$ for $V=10$ and $\delta=0.2$ in Fig.~\ref{fig:p_up_down}(b). The figure depicts that initially for $U=0$, $P_{\uparrow\uparrow}\sim 1$. As $U$ increases and becomes $U\sim V = 10$, $P_{\uparrow\uparrow}$ reaches a minimum value close to zero and at the same time $P_{\uparrow\downarrow}$ becomes maximum, indicating the dissociation of the NN pair ($\uparrow\uparrow$) and formation of a stable doublon. Further increase in the value of $U$ leads to the stability of the NN pair ($\uparrow\uparrow$) state as the energy of the doublon state is off resonant to the NN pair ($\uparrow\uparrow$) state which prohibits an NN pair ($\uparrow\uparrow$) breaking - a situation similar to the case when $U < V$.

We further examine the behaviour of the probability of doublon formation as a function of $\delta$ by plotting the saturated values of $P_{\uparrow\downarrow}$ (filled circles) and $P_{\uparrow\uparrow}$ (empty circles) in Fig.~\ref{fig:p_up_down}(c) for $U=V=10$ at $t=10(J^{-1})$. 
For $\delta=0$, the values of $P_{\uparrow\downarrow} \sim 1$ and $P_{\uparrow\uparrow}\sim 0$ are the clear indication of a stable doublon formation. However, as $\delta$ increases, $P_{\uparrow\downarrow}$ decreases and $P_{\uparrow\uparrow}$ increases and at $\delta\sim 1$, we have almost equal probabilities of both the bound states. Further increase in $\delta$ (i.e. for $\delta > 1$) results in further decrease in $P_{\uparrow\downarrow}$ and increase in $P_{\uparrow\uparrow}$ indicating that the  NN pair ($\uparrow\uparrow$) is favourable at higher hopping imbalance.

This analysis suggest that a stable doublon can be formed in the QW of two-component fermions by suitably tuning the onsite as well as NN interaction along with the hopping imbalance. However, observation of such stable doublons is challenging using the quantum simulators like ultracold atoms in optical lattices due to the complexities involved. In such a circumstance, quantum computing simulation can provide a platform to observe such phenomenon using existing NISQ devices. In the rest of the paper we mainly focus on to observe the signatures of such non-trivial doublon formation and its stability in quantum computing simulation. 

{\em Signatures from quantum circuit simulations.-} Before proceeding further we briefly provide the quantum circuit implementation of the QW considered here. 
The QW is simulated using the $ibmq\_brisbane$ and $IonQ$-$Aria~1$ quantum hardware. The circuit for the two-component model shown in Eq.~\ref{eqn:ham} is constructed by considering $7$ qubits for each component which requires $14$ qubits in total. The required initial state for the time evolution is constructed by applying the Pauli-X gate to the default initial state. The quantum circuit for the unitary time evolution is obtained using the Suzuki-Trotter~\cite{SUZUKI1990319} decomposition of the operator as
\begin{equation}
    \hat{U}(t)=e^{-i\hat{H}t}=(e^{-i\hat{H}\Delta t})^n
\end{equation}
with the time step $\Delta t=\frac{t}{n}$ (see supplementary materials~\cite{supplementary} for  details). The time evolved state $|\psi(t)\rangle$ at $t=n\Delta t$  is obtained by applying 
$\hat{U}(t)$ on an initial state with $n$ Trotter steps and considering a large $n$ to reduce the local and global errors. To avoid the variable circuit depth for each time step, we convert our Trotter circuit of time evolution operator $e^{-i\hat{H}t}$ to a constant depth parametrized circuit~\cite{Khatri2019quantumassisted, Jones2022robustquantum, heya2018variational}, using the QUIMB library~\cite{quimb}.

The parametrized circuit is shown in Fig.~\ref{fig:toy_plot}(c), which consists of alternating layers of single-qubit $U_3$ gates and two-qubit control-Z gate. Here the $U_3$ gate consists of three rotational parameters and is defined as,
\begin{equation}
U_3(\theta, \phi, \lambda)=
    \begin{pmatrix}
        \cos(\frac{\theta}{2}) & -e^{i\lambda}\sin(\frac{\theta}{2}) \\
        e^{i\phi}\sin(\frac{\theta}{2}) & e^{i(\phi+\lambda)}\cos(\frac{\theta}{2}).
    \end{pmatrix}
\end{equation}
%In our case, we consider a circuit consisting of nine $U_3$ layers and eight two-qubit gate layers.
Denoting the parametrized circuit as an unitary operator $\hat{\mathcal{U}}^*(\Theta)$, with  $\Theta$ as all the rotation angles of the $U_3$ gates and denoting the Totter circuit by $\hat{\mathcal{U}}$ and the initial state by $|\psi(0)\rangle$, the optimal recompiled unitary that mimics the Trotter circuit is obtained by maximizing the Fidelity 
\begin{equation}
    F(\Theta) = \Big{|}\langle \psi(0)|\hat{\mathcal{U}}^*(\Theta)^\dagger \hat{\mathcal{U}}|\psi(0)\rangle\Big{|}^2.
\end{equation} 
Following the above method, although we are able to construct a shallower depth circuit, due to the noise in the device we still get some undesirable states in the output. To circumvent this we implement the post-selection (PS)~\cite{Smith2019, McArdle2019} and zero noise extrapolation (ZNE)  error mitigation methods ~\cite{ZNE_Li_2017, ZNE_Temme_2017, ZNE_Kandala2019}(see supplementary materials~\cite{supplementary} for details), which significantly increases the accuracy of the results. For all our calculations on $ibmq$ and $IonQ$ quantum hardware, we use $6000$ and $5000$ shots respectively.

With this setup in hand, we realize our numerical prediction of doublon formation by implementing the system Hamiltonian on different quantum processors for a small system.
As already predicted in the classical simulation, a small value of $\delta$ is preferable for stable doublon formation, we choose $\delta = 0.2$ and fix the resonance condition for interaction i.e. $U=V=10$ for the time evolution.  Fig.~\ref{fig:p_up_down}(d) shows the data for $P_{\uparrow\uparrow}$ (empty symbols) and $P_{\uparrow\downarrow}$ (solid symbols) as a function of $t(J_{\uparrow}^{-1})$ without any error mitigation or noisy data (blue triangle), with PS (black circle) and with  PS+ZNE (red square) error mitigation, which are computed using the $ibmq\_brisbane$ hardware. Besides the $ibmq\_brisbane$ data we also show the PS error mitigated data  obtained from the $IonQ$-$Aria~1$ hardware (magenta cross in Fig.~\ref{fig:p_up_down}(d)). For comparison, we compute the relevant quantities using the ED method for an equivalent system size. The data without error-mitigation shows qualitative agreement with the exact result (green star). However, the error-mitigated results show a good quantitative agreement with the ED  results. Although both $ibmq$ and $IonQ$ error mitigated data show qualitatively similar results with the results obtained from ED, the rest of our calculations are done using $ibmq$ only. In Fig.~\ref{fig:p_up_down}(e) we  plot $P_{\uparrow\downarrow}$ and $P_{\uparrow\uparrow}$ as a function of $U$ at a particular time $t=1.6(J^{-1})$ of the time evolution, for $V=10$ and $\delta = 0.2$. This clearly establishes the resonance condition for doublon formation as predicted from the TEBD analysis. Finally similar agreement is also seen in Fig.~\ref{fig:p_up_down}(f) for  $P_{\uparrow\downarrow}$ and $P_{\uparrow\uparrow}$ when plotted against $\delta$ at time $t=1.6{J^{-1}}$ for $V=U$ (compare with Fig.~\ref{fig:p_up_down}(c)). 
\begin{figure}[t!]
    \centering
    \includegraphics[width=1\columnwidth]{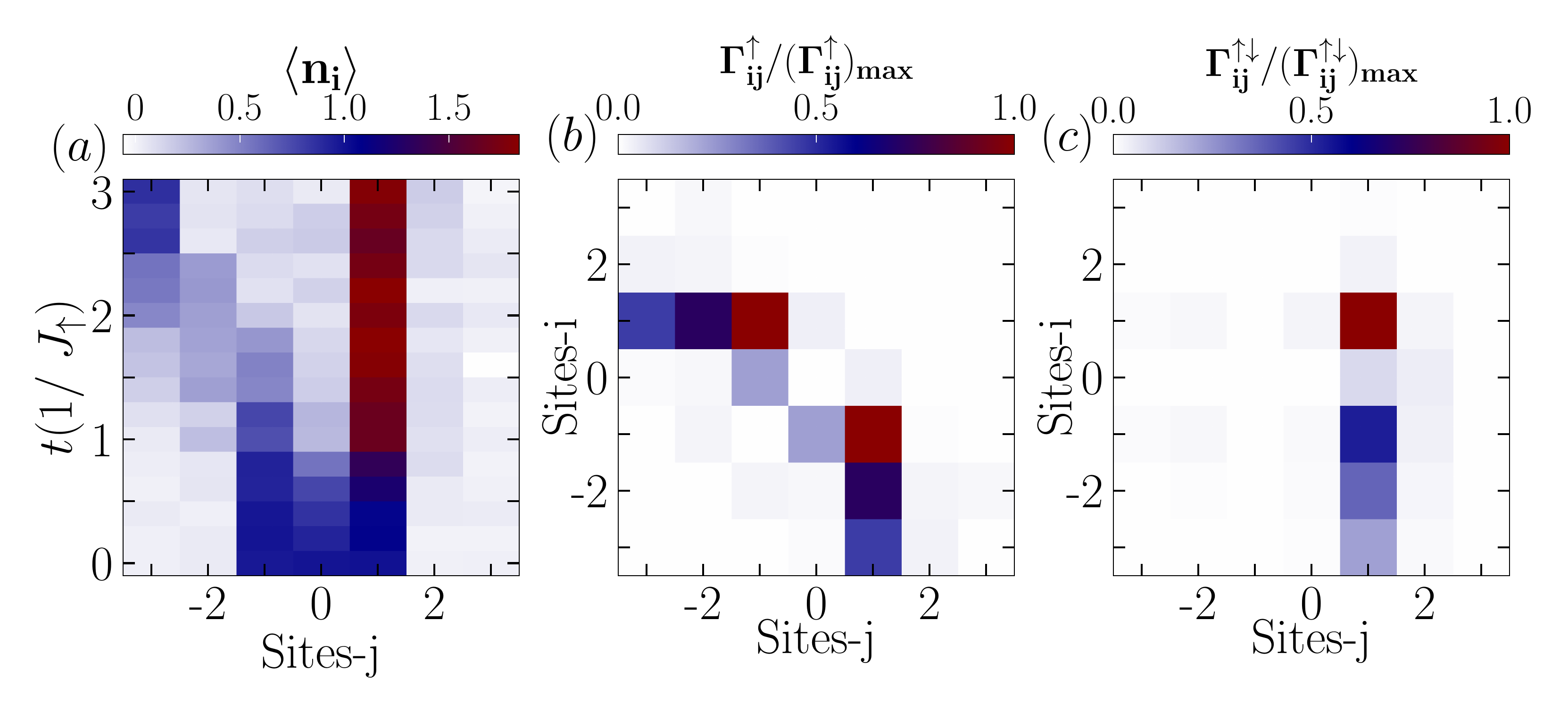}
    \caption{(a), (b) and (c) show the PS and ZNE error mitigated data of the time evolved density $\langle n_i\rangle$, correlation function  $\Gamma_{ij}^{\uparrow}$ and $\Gamma_{ij}^{\uparrow\downarrow}$ respectively, for $U=V=10$ and $\delta=0.2$ obtained using the $ibmq\_brisbane$ hardware. The correlation functions are plotted at time $t=1.6(J^{-1})$ of the time evolution.}
    \label{fig:den_evo}
\end{figure}

To further quantify the above observations, we plot the total particle density $\langle \hat{n}_i\rangle = \langle \hat{n}_{i,\uparrow}\rangle + \langle \hat{n}_{i,\downarrow}\rangle$ for each site as a function of time shown in Fig.~\ref{fig:den_evo}(a), for fixed hopping imbalance $\delta=0.2$ and interaction strength $V=U=10$. The information about the doublon formation can be seen from the value $\langle \hat{n}_i \rangle \sim 2$ at the $1^{st}$ site after a short time evolution and the independent  dynamics of the $\uparrow$ particle on the left part of the central site as depicted in Fig.~\ref{fig:den_evo}(a). Additionally, we also calculate the inter- and intra-component density density correlation
\begin{equation}
      \Gamma_{ij}^{\uparrow\downarrow}(t)=\langle \hat{n}_{i,\uparrow} \hat{n}_{j,\downarrow}\rangle \ \ and \ \ \Gamma_{ij}^{\sigma}(t)=\langle \hat{a}_{i, \sigma}^\dagger  \hat{a}_{j, \sigma}^\dagger \hat{a}_{j, \sigma}\hat{a}_{i, \sigma} \rangle
      \label{eqn:corr}
\end{equation}
to identify the type of bound states present in the system. $\Gamma_{ij}^{\uparrow}$ and $\Gamma_{ij}^{\uparrow\downarrow}$ are plotted at time $t=1.6(J^{-1})$ in Fig.~\ref{fig:den_evo}(b) and (c) respectively for same parameter values as in Fig.~\ref{fig:den_evo}(a). The presence of a diagonal element in the $\Gamma_{ij}^{\uparrow\downarrow}$ as shown in Fig.~\ref{fig:den_evo}(c) indicates the formation of an onsite doublon in the system. The weaker values of $\Gamma_{ij}^{\uparrow}$ immediately above or below the diagonal in Fig.~\ref{fig:den_evo}(b) indicate the lesser probabilities of finding an NN  pair ($\uparrow\uparrow$).
For all the calculations in Fig.~\ref{fig:p_up_down} and ~\ref{fig:den_evo} we consider the parametrized circuit with eight layers of alternative single qubit $U_3$ and two-qubit control-Z gate. Note that for most parameter values considered here we find the fidelity $F(\Theta)$ is greater than $99\%$ and in fewer cases it is below $99\%$.

With this observation we confirm the dynamical creation of stable doublons in the QW of interacting fermions. Now it is essential to examine the stability of such a doublon in the dynamics which will be discussed in the following section. 
\begin{figure}[t!]
    \centering
    \includegraphics[width=1\columnwidth]{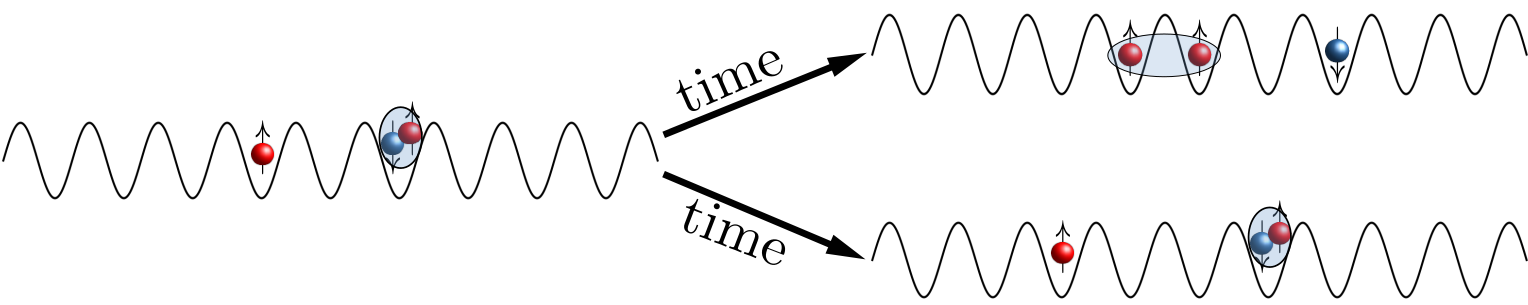}
    \caption{ Shows the schematic for two energy-conserving time evolution processes when we start from the initial state $|\psi(0)\rangle=\hat{a}_{-1,\uparrow}^{\dagger} \hat{a}_{1,\uparrow}^{\dagger}\hat{a}_{1,\downarrow}^{\dagger}|vac\rangle$.}
    \label{fig:toy_dissociation}
\end{figure}

\begin{figure}[t!]
    \centering
    \includegraphics[width=1\columnwidth]{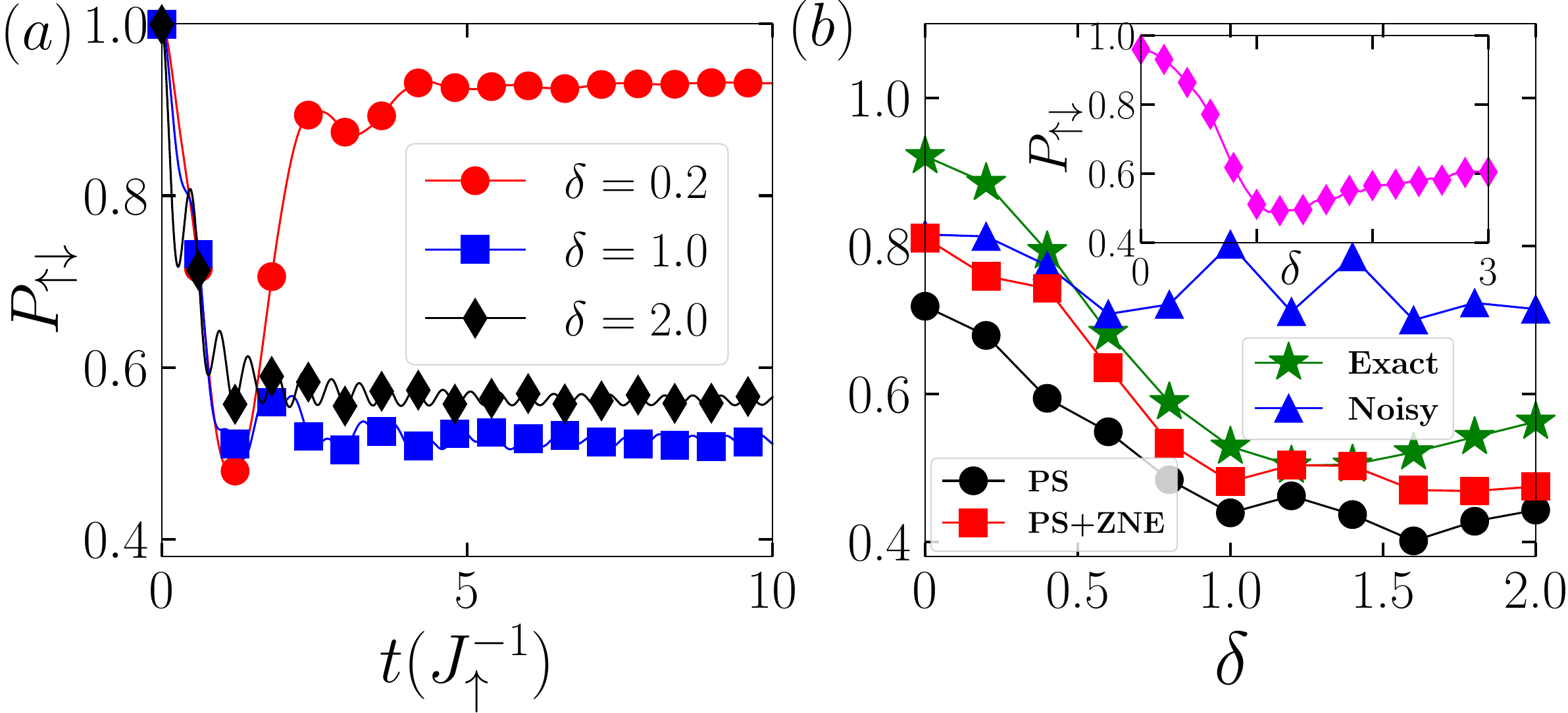}
    \caption{(a) Shows the TEBD data for $P_{\uparrow\downarrow}$ as a function of $t(J_{\uparrow}^{-1})$ for different values of $\delta$ when $U=V=10$ and $L=51$. (b) Shows the digital quantum simulation data for  $P_{\uparrow\downarrow}$ plotted as a function of $\delta$ at time $t=2.5(J_{\uparrow}^{-1})$ for $U=V=10$. The green stars, bule triangles, black circles and red squares denote the exact, noisy, PS and PS+ZNE data respectively obtained using the $ibmq\_brisbane$ hardware. Inset shows $P_{\uparrow\downarrow}$ at time $t=10(J_{\uparrow}^{-1})$ as a function of $\delta$ obtained using TEBD for $U=V=10$ and $L=51$.}
    \label{fig:dissociation}
\end{figure}

{\em Stability of doublon.-} To examine the stability of the doublon we consider an initial state with one doublon and one $\uparrow$ particle at the right and left sites of the central site respectively. Such initial state which is given as $|\psi(0)\rangle=\hat{a}_{-1,\uparrow}^{\dagger} \hat{a}_{1,\uparrow}^{\dagger}\hat{a}_{1,\downarrow}^{\dagger}|vac\rangle$ and as depicted in Fig.~\ref{fig:toy_dissociation} (left) guarantees a maximum probability of doublon dissociation. Naively, one would expect that for larger values of $\delta$, the onsite doublon will tend to dissociate completely and a stable  NN pair ($\uparrow\uparrow$) will be formed as the latter one is energetically more stable. However, here we obtain a counter-intuitive situation. From the TEBD data shown in Fig.~\ref{fig:dissociation}(a), we can see that $P_{\uparrow\downarrow}$ after some time saturates to finite values for different values of $\delta$ when $U=V=10$. As expected, for $\delta=0.2$, $P_{\uparrow\downarrow}$ saturates to a larger value close to one (red circles). For $\delta=1$, $P_{\uparrow\downarrow}$ saturates to a value close to $0.5$ (blue squares) due to the equal probability of the doublon and the NN pair ($\uparrow\uparrow$) states. However, when $\delta$ becomes larger than one, we obtain that  $P_{\uparrow\downarrow}$  saturates to larger values compared to that for $\delta=1$ which is shown as black diamonds for $\delta=2$.

We also perform quantum computing implementation to observe the stability of the initially created doublon as a function of $\delta$. 
We plot the value of $P_{\uparrow\downarrow}$ at time $t=2.5(J_{\uparrow}^{-1})$ as a function of $\delta$ in Fig.~\ref{fig:dissociation}(b) for $U=V=10$ where the data from the ED and quantum computing simulation show the initial decrease and then increase of $P_{\uparrow\downarrow}$ with increase in $\delta$. The inset of Fig.~\ref{fig:dissociation}(b) shows the trend obtained from the TEBD simulations up to $\delta=3$ which confirms the results obtained from ED. The discrepancy between the ED and quantum computing simulation in Fig.~\ref{fig:dissociation}(b) is due to the increase in noise with increase in circuit depth. 
Here we have considered a parametrized circuit of $12$ control-Z layers. For most of the parameter values we get the fidelity $F(\Theta)>97\%$ and in fewer cases we get lower fidelity than this. 

This analysis shows that once the doublon is formed it remains stable at least up to 50 percent probability and never vanishes. Surprisingly, after reaching a minimum, the probability of stable doublon increases with increase in the hopping imbalance at the resonance condition i.e. $U=V$.

{\em Conclusions.-}
In this work, we observed the signature of an onsite doublon formation in the dynamics of interacting fermions using NISQ devices. By implementing appropriate digital circuits for the Fermi-Hubbard model and utilizing the circuit optimization and various error mitigation techniques we obtained that the onsite doublons are spontaneously formed in the dynamics due to the combined effect of two competing interactions such as the inter-particle onsite interaction and the intra-particle NN interaction. The probability of doublon formation is found to be strongly dependent on the hopping imbalance between the constituents. Furthermore, we  examined the stability of such doublons by analyzing the QW from an initial state with already formed doublon and a free particle where we obtained that in any circumstances the probability of doublon survival is higher than the probability of doublon dissociation.

Our work reveals two important results: (i) a route to form a stable doublon in the QW of non-local fermions due to interaction. (ii) signature of such non-trivial doublon in a quantum computing simulation which was not observed in any other quantum simulators. This also opens up avenues and provides appropriate platform to explore dynamics of more interacting particles and their combined dynamics. One immediate extension can be the formation of two or more doublons in the dynamics and their stability against external perturbations such as disorder and tilt. 

{\em Acknowledgement.-}
We thank Sudhindu Bikash Mandal for useful discussions. T.M. acknowledges support from Science and Engineering Research Board (SERB), Govt. of India, through project No. MTR/2022/000382 and STR/2022/000023. T.M. also acknowledges MeitY QCAL project for providing access to use IonQ hardware through AWS Braket platform.

\bibliography{reference_new}
\clearpage
\onecolumngrid
\begin{center}
\textbf{Supplementary materials for 
 ``Realizing non-trivial doublon formation using a quantum computer"}
\end{center}
In this supplemental material we discuss the circuit implementation of the model considered and various error mitigation techniques used for the simulations.

\twocolumngrid
\subsection{Jordan-Wigner transformation}
For the implementation of our operator on the quantum circuit we need to transform our Hamiltonian on spin-$\frac{1}{2}$ basis. We use the Jordan-Wigner transformation~\cite{JWT_somma_2002} to map our Hamiltonian from Fock-space basis to spin-$\frac{1}{2}$ basis. The transformations of the operators are given as,
\begin{equation*}
    \hat{a}_{j, \uparrow} =\left( \prod_{i=0}^{j-1}\hat{\sigma}_i^{z}\right)\frac{\hat{\sigma}_{j}^x-i\hat{\sigma}_{j}^y}{2}, \ \ \ 
    \hat{n}_{j, \uparrow}  = \frac{1}{2}(1-\hat{\sigma}_{j}^z)
\end{equation*}
\begin{equation*}
    \hat{a}_{j, \downarrow} =\left( \prod_{i=L}^{L+j-1}\hat{\sigma}_i^{z}\right)\frac{\hat{\sigma}_{L+j}^x-i\hat{\sigma}_{L+j}^y}{2}, \ \ \ 
    \hat{n}_{j,\downarrow}  = \frac{1}{2}(1-\hat{\sigma}_{L+j}^z),
\end{equation*}
where $\hat{\sigma}^{x}, \hat{\sigma}^{y}$ and $\hat{\sigma}^{z}$ are the Pauli matrices. One can check that the transformation preserves anti-commutation relation i.e. $\{\hat{a}_{i,\sigma}, \hat{a}_{j,\sigma}^{\dagger}\}=\delta_{ij}$, $\{\hat{a}_{i,\sigma}^{\dagger}, \hat{a}_{j,\sigma}^{\dagger}\}=0$ and $\{\hat{a}_{i,\sigma}, \hat{a}_{j,\sigma}\}=0$. We consider a one-dimensional system with $L$ lattice sites. As we are considering two different component systems we require $2L$ qubits to simulate our system. Here, we assign first $L$ qubits for $\uparrow$ component particles and remaining $L$ sites for the $\downarrow$ component particles.

With this, the transformed Hamiltonian corresponding to Eq.~1 in the main text is given as,
\begin{equation}
    \hat{H} = \hat{H}_0+\hat{H}_1+\hat{H}_2,
    \label{eqn:qubit_ham}
\end{equation}
where,
\begin{flalign*}
    & \hat{H}_0 = \sum_{j=0}^{L-2}\frac{J_{\uparrow}}{2}(\hat{\sigma}_{j}^{x}\hat{\sigma}_{j+1}^{x}+\hat{\sigma}_{j}^{y}\hat{\sigma}_{j+1}^{y})+\frac{V_{\uparrow\uparrow}}{4} \hat{\sigma}_{j}^{z}\hat{\sigma}_{j+1}^z \\
    &+\sum_{j=L}^{2L-2}\frac{J_{\downarrow}}{2}(\hat{\sigma}_{L+j}^{x}\hat{\sigma}_{L+j+1}^{x}+\hat{\sigma}_{L+j}^{y}\hat{\sigma}_{L+j+1}^{y})
    +\frac{V_{\downarrow\downarrow}}{4} \hat{\sigma}_{L+j}^{z}\hat{\sigma}_{L+j+1}^z \\
    & \hat{H}_1 = \frac{U_{\uparrow\downarrow}}{4}\sum_{0}^{L-1}\hat{\sigma}_j^z\hat{\sigma}_{L+j}^z\\
    & \hat{H}_2 = \sum_{j=1}^{L-2}(-\frac{V_{\uparrow\uparrow}}{2}\hat{\sigma}_{j}^{z}-\frac{V_{\downarrow\downarrow}}{2}\hat{\sigma}_{L+j}^{z})-\frac{V_{\uparrow\uparrow}}{4}(\hat{\sigma}_0^z+\hat{\sigma}_{L-1}^{z})\\
    &-\frac{V_{\downarrow\downarrow}}{4}(\hat{\sigma}_L^z+\hat{\sigma}_{2L-1}^{z})
\end{flalign*}
Here we consider $V_{\uparrow\downarrow}=0$ as mentioned in the main text. We use this Hamiltonian (Eq.~\ref{eqn:qubit_ham}) to write the time evolution operator $\hat{U}$ of the main text. This unitary time evolution operator $\hat{U}$ can be converted into the quantum circuit and the details of which is given in the next section.

\subsection{Time evolution on Quantum Circuit}
%To probe the dynamical properties of our few particle systems we use the CTQW, which is nothing but the time evolution of our few particle systems governed by our Hamiltonian $H$. 
\begin{figure}[t!]
    \centering
    \includegraphics[width=1\columnwidth]{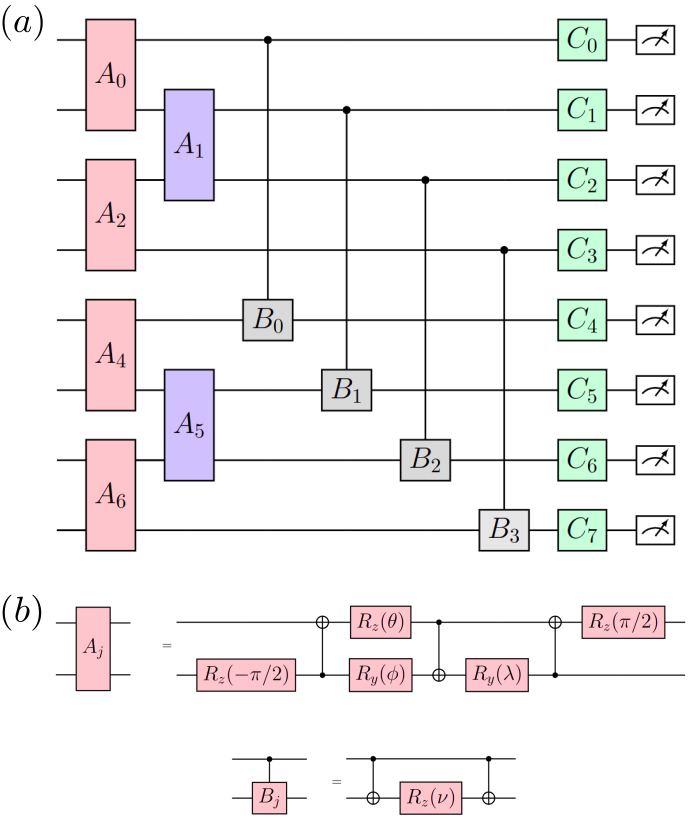}
    \caption{(a)Shows the circuit diagram for a single Trotter step. (b) Shows the explicit circuit diagram for the operators $A_{j}$ and $B_j$. Here we use $\theta=\frac{\pi}{2}-\frac{V_{\sigma\sigma}\Delta t}{2}$, $\phi=J_\sigma\Delta t-\frac{\pi}{2}$, $\lambda=\frac{\pi}{2}-J_\sigma\Delta t$ and $\nu=\frac{U_{\uparrow\downarrow}\Delta t}{2}$.}
    \label{fig:trotter_ckt}
\end{figure}
% The time evolution is described by schr\"{o}dinger equation,
% \begin{equation}
%     \frac{d|\psi(t)\rangle}{dt}=-i\hat{H}|\psi(t)\rangle
% \end{equation}
% Here we consider $\hbar=1$ and  $\hat{H}$ is the system Hamiltonian. For a given initial state $|\psi(0)\rangle$, the time-evolved quantum state at a time $t$ is given by $|\psi(t)\rangle = e^{-i\hat{H}t}|\psi(0)\rangle$. To probe the dynamical properties we implement the time evolution operator $\hat{U}=e^{-i\hat{H}t}$ on a quantum circuit. 
% Decomposing the qubit system Hamiltonian is not easy, because the qubit strings summed over may not commute with each other, i.e. we can't simply write $e^{\hat{X}+\hat{Y}}\neq e^{\hat{X}}e^{\hat{Y}}$ if the two operators $\hat{X}$ and $\hat{Y}$ don't commute with each other. The solution to this problem comes in terms of the Suzuki-Trotter~\cite{SUZUKI1990319} decomposition at the cost of a certain amount of error. 
The first order Suzuki-Trotter decomposition of our time evolution operator $\hat{U}(t)$ for $n$ number of Trotter steps is given by,
\begin{align*}
    \hat{U}(t) &= e^{-i\hat{H}t}= \Big(e^{-i\hat{H}\Delta t}\Big)^n \\
    &= \Big(e^{-i\hat{H_0}\Delta t}e^{-i\hat{H_1}\Delta t}e^{-i\hat{H_2}\Delta t}\Big)^n + \mathcal{O}(n\Delta t^2)\\
&=\Big(\prod_{j=even}\hat{A}_{j}\prod_{i=odd}\hat{A}_{j}\prod_{j=0}^{L-1}\hat{B}_{j}\prod_{j=0}^{2L-1}\hat{C}_{j}\Big)^{n}+\mathcal{O}(n\Delta t^2).
\end{align*}
To make the notation simpler we use,
\begin{align*}
    \hat{A}_j&=e^{-i\Delta t[J_{\sigma}/2(\hat{\sigma}_{j}^{x}\hat{\sigma}_{j+1}^{x}+\hat{\sigma}_{j}^{y}\hat{\sigma}_{j+1}^{y})+(V_{\sigma}/4) \hat{\sigma}_{j}^{z}\hat{\sigma}_{j+1}^z]},\\
    \hat{B}_j&=e^{-i\Delta t(U_{\uparrow\downarrow}/4)\hat{\sigma}_j^z\hat{\sigma}_{L+j}^z},\\
    \hat{C}_j &= e^{i\Phi_j\hat{\sigma}_j^z}
\end{align*}

In the case of $\hat{A}_j$, $j$ can take values as $0\leq j \leq L-2$ and $L\leq j \leq 2L-2$, for $B_j$, $j$  runs from $0$ to $L-1$ and for $C_j$, $j$ can take values from $0$ to $2L-1$.  Depending upon the sites $j$, the onsite rotation angles $\Phi_j$($\frac{V_{\uparrow\uparrow}\Delta t}{2}$ or $\frac{V_{\downarrow\downarrow}\Delta t}{2}$ or $\frac{V_{\uparrow\uparrow}\Delta t}{4}$ or $\frac{V_{\downarrow\downarrow}\Delta t}{4}$) changes. The Fig.~\ref{fig:trotter_ckt}(b) shows the quantum circuit for the expression $\hat{A}_j$ with the optimal number of gates~\cite{vatan2004,Smith2019}. The first-order Trotter circuit for a single Trotter step is shown in Fig.~\ref{fig:trotter_ckt}(a).  The error depends on the time step $\Delta t$, which can be decreased by making the step size smaller. However, this will require more number Trotter steps which in turn increases the number of noisy quantum gates in the circuit. Therefore, we consider a moderately small $\Delta t=0.1$ in our calculation.

\section{Error Mitigation}
To construct the quantum circuit we use the Trotterization method. However, the problem with this method is the number of noisy quantum gates increases with time steps. We use the circuit recompilation(a brief discussion given in the main text) method to reduce the circuit depth but still a sufficient number of noisy quantum gates appear in our circuit which reduces the accuracy of the results. We use the following error mitigation method to increase the accuracy of our results.

\subsection{Post-Selection Method}

The post-selection method is suitable for increasing the accuracy of the results by respecting the symmetry of the Hamiltonian. Suppose $\hat{H}$ is our system Hamiltonian and $\hat{O}$ is some measurable observable in our system and is a conserved quantity i.e. $[\hat{H}, \hat{O}]=0$. Therefore, the expectation value of the observable $\hat{O}$ does not change as a function of time, the expectation values of $\hat{O}$ at time $t=0$ should be equal to the expectation value of $\hat{O}$ after time $t=t_f$. The exact simulation satisfies the condition $\langle \hat{O}\rangle_{t=0}=\langle \hat{O}\rangle_{t=t_f}$, but in case of noisy simulation some undesirable state appears which does not respect conservation rule. After measurement, these unphysical states may be discarded.

%The conserved quantities depend upon the Hamiltonian of the system like if we consider some isolated system the particle number will be conserved, but if the system is governed by a quadratic Hamiltonian then we discard the incorrect parity. In our case, we conserve the number of particles.

The Hamiltonian we consider conserves the number of particles of each component. As our system contains two different component particles with a specific numbers, we consider those output states for which $N_\uparrow=\sum_j\langle \hat{a}_{j,\uparrow}^{\dagger}\hat{a}_{j,\uparrow}\rangle=2$ and $N_\downarrow=\sum_j\langle \hat{a}_{j,\downarrow}^{\dagger}\hat{a}_{j,\downarrow}\rangle=1$. We take $N_\uparrow$ and $N_\downarrow$ as post-selection for our system. For calculation, we consider counts of those states that satisfy the above two relations for other states we consider counts to be zero.

\subsection{Zero Noise Extrapolation Method}
ZNE is a well-established method for improving computing results by mitigating incoherent noise. The practical implementation of this technique is very suitable for the near-term hardware. The idea is based on extrapolating the expectation values of some observable in the noiseless limit by calculating the expectation values of the same observable in different noise limits. This procedure has two major steps $(i)$ scale up the circuit noise level and $(ii)$ extrapolate the noisy expectation values for zero noise.\\
\begin{figure}[t!]
    \centering
    \includegraphics[width=1\columnwidth]{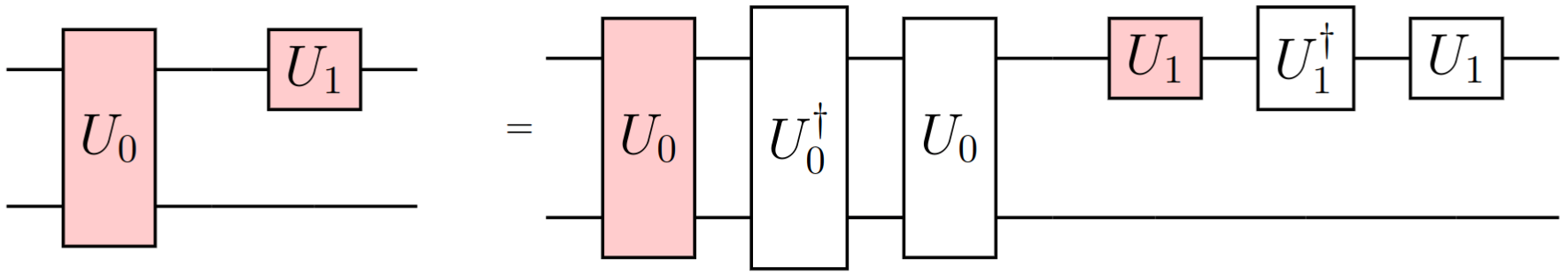}
    \caption{Shows a unitary circuit with gate-level local folding for noise scaling $\lambda=3$.}
    \label{fig:ZNE_ckt}
\end{figure}
$(i)$ Scale up the circuit noise: Scaling up the noise in quantum circuits can be done using different methods. Pulse-stretching~\cite{ZNE_Temme_2017} and gate-level unitary folding~\cite{ZNE_Li_2017} are two methods for scaling up the noise in the circuits. Point to note that the expectation values of the observable don't change for pulse-stretching and gate-level unitary folding if we use a noiseless system. As the real quantum hardware are noisy system we observe the output expectation values worsen if the noise scales up. For our calculations, we go with the gate-level unitary folding method to scale up the noise. A toy circuit diagram for local folding is shown in Fig.~\ref{fig:ZNE_ckt}. We utilize this local unitary operation to scale up the noise, employing the open-source library mitiq~\cite{mitiq}. Other types of gate-level folding can be found in Ref~\cite{Giurgica-Tiron2020}. Let $\eta$ be the noise level of the original circuit and the noise of the scaled circuit defined as $\eta^\prime=\lambda\eta$. For our extrapolation calculation, we run our circuit for three different noise scales $\lambda=1$(original circuit), $\lambda=2$(doubling the circuit depth), and $\lambda=3$(increasing the circuit depth threefold). As the two-qubit gates are much more noisy than the single-qubit gates, we consider only two-qubit gates for noise scaling. After getting the noisy expectation values we extrapolate it for zero noise. The extrapolation technique is given as follows.\\

$(ii)$ Extrapolation of noisy expectation values: Let, $\hat{O}$  be the observable we want to calculate. The expectation values of the observable $O_\lambda$ for noise scale $\eta^\prime = \lambda\eta$. Once $\lambda$ and $O_\lambda$ values are known we can easily calculate zero noise extrapolated value of the observable($O_{\lambda=0}$). For extrapolating into the zero noise we use Richardson's extrapolation~\cite{ZNE_Temme_2017} technique.

\end{document}